# Electrical characteristics of vertical-geometry Schottky junction to magnetic insulator (Ga,Mn)N heteroepitaxially grown on sapphire

Karolina Kalbarczyk[1*], Krzysztof Dybko[1,2], Katarzyna Gas[1,3], Dariusz Sztenkiel[1], Marek Foltyn[1], Magdalena Majewicz[1], Piotr Nowicki[1], Elżbieta Łusakowska[1], Detlef Hommel[3,4], and Maciej Sawicki[1*]

[1]*Institute of Physics, Polish Academy of Sciences, Aleja Lotnikow 32/46, PL-02668 Warsaw, Poland*

[2]*International Research Centre MagTop, Institute of Physics, Polish Academy of Sciences, Aleja Lotnikow 32/46, PL-02668 Warsaw, Poland*

[3]*Institute of Experimental Physics, University of Wrocław, Pl. Maxa Borna 9, 50-204 Wrocław, Poland*

[4]*Polish Center of Technology Development, ul. Stabłowicka 147, 54-066, Wrocław, Poland*

*E-mail: kalbarczyk@ifpan.edu.pl,   mikes@ifpan.edu.pl

Schottky barrier height and the ideality factor $\eta$ are established for the first time in the single phase (Ga,Mn)N using a vertical geometry device. The material has been heteroepitaxially grown on commercially available low threading dislocation density GaN:Si template. The observed above 10 MΩ resistances already at room temperature are indicative that a nearly conductive-dislocation-free electrical properties are achieved. The analysis of temperature dependence of the forward bias I-V characteristics in the frame of the thermionic emission model yields Ti-(Ga,Mn)N Schottky barrier height to be slightly lower but close in character to other metal/GaN junctions. However, the large magnitudes of the ideality factor $\eta > 1.5$ for T ≤ 300 K, point to a sizable current blocking in the structure. While it remains to be seen whether it is due to the presence of (Ga,Mn)N barrier or due to other factors which reduce the effective area of the junction, an existence of a substantial serial resistance may hold the key to explain similar observations in other devices of a corresponding structure and technological relevance.

# 1. Introduction

Besides becoming the material system of choice for nowadays optoelectronic [1] and high-power applications [2], GaN-based and related alloys and heterostructures are expected also to play a major role in the realization of spin-related functionalities based on semiconductors [3][4][5]. While the search for a technology-viable nitride semiconductor exhibiting (ferro-)magnetic properties at room temperature is still the subject of active research, a great deal of knowledge on the underlying physical processes has been accumulated from the investigation of these system at their relevant temperatures [6][7][8]. In the line of our research is (Ga,Mn)N – an emerging ferromagnetic semiconductor whose long range ferromagnetic ordering has been confirmed at the low end of cryogenic temperatures [9][10]. This magnetic form of GaN, in which a small percentage of Ga atoms is randomly substituted by Mn, belongs to increasingly important group of Rashba materials [3]. The established strong mid-gap Fermi-level pinning [11][12] and the absence of a depletion layer, in combination with its insulating character [13][14][15][16] and a sizable dielectric strength of at least 5 MV/cm [17], points at Mn containing GaN as a worthwhile insulating buffer material for applications in (high power) nitride devices. By the same token this ferromagnetic insulator appears to be an ideal building block for nitride based spin harnessing concepts, which could further stimulate the development of all-nitride low power information processing and wireless communication technologies based on spin waves, the so called magnonics [18]. On the other hand, in the field of spintronics, an experimental verification of theoretical predictions suggesting large spin filtering capabilities of spin filters [19] and resonant tunneling devices [20] with (Ga,Mn)N as the barrier material is still lacking. The sought functionality is provided here by the spin-splitting of (Ga,Mn)N conduction band that creates the spin-dependent barriers, which should give rise to a highly spin-polarized current.

However, the observation of the last effect proved futile in the GaN based structures [21] due to the presence of a large number of conducting dislocations [22][23], which provide a short-circuiting path across the (Ga,Mn)N barrier layer. Concerning the nitride structures grown on GaN templates, as in the case of the present study, pure screw threading dislocations had been identified as the primary source of this detrimental leakage [24]. The presence of the same conductive dislocations hinders also the development of such important concepts as all-nitride high electron mobility transistor (HEMT), or heterostructure field-effect transistor (HFET) [25][26][27][28].

In this study we experimentally investigate and numerically quantify the temperature dependence of the Schottky barrier height formed at a junction of Ti/Au metallic contact to (Ga,Mn)N. The investigated junction is prepared in the vertical geometry. The junction itself has been defined by an opening in the insulating $Al_2O_3$ layer which constitutes also the base for a macroscopically large top electrical contact. The observed very large resistance of the junction indicates that a nearly conductive-dislocation-free electrical properties are achieved. We show that the devices fabricated by this approach can be characterized by typical magnitudes of the metal-nitride semiconductor barrier heights, however some deviations from the thermionic emission model are clearly observed. Since the structural resemblance of our devices to HEMT or similar structures, the findings reported here may prove useful for that ever so much important field of material science.

## 2. Experimental details

The investigated (Ga,Mn)N layer with Mn concentration of about 5% is grown by plasma assisted molecular beam epitaxy (MBE) method in a Scienta-Omicron Pro-100 MBE at 620°C under near stoichiometric conditions [29]. We use a *c*-plane sapphire based commercially available low threading dislocation density conductive n-type (0001) GaN:Si template, which contains, generally termed, $SiN_x$ mask [30]. According to the technical specification the Si concentration in the n-type GaN buffer layer amounts to $3 \times 10^{18}$ cm$^{-3}$, so the buffer layer serves also as the conducting bottom contact in the structure. This highly advantageous feature of the substrate for the design of vertical devices, has got a certain drawback as it precludes a direct electrical testing of the (Ga,Mn)N layer. However, much can be learnt from the magnetic investigation. So, despite that the electrical studies presented here are undertaken only down to 150 K, we perform the magnetic characterization at low temperatures to infer the electrical properties of the (Ga,Mn)N layer.

(Ga,Mn)N grown at identical conditions as in ref. [29] undergoes the Curie transition between 5 and 8 K. This so far exclusively low temperature feature is provided by ferromagnetic superexchange interaction between $Mn^{3+}$ ions in GaN [31][32][33][34]. This prerequisite condition for homogenous ferromagnetism in diluted (Ga,Mn)N [35] has been confirmed by X-ray absorption spectra near the Mn *K* absorption edge reported in refs. [36][37], and, most importantly here, in ref. [29], where also other relevant characterization details are given. Therefore, an observation of (ferro-)magnetic characteristics in (Ga,Mn)N as those reported in e.g. refs. [17][36] can be taken as an indirect indication of the dominance of the $Mn^{3+}$ state in the material, largely reducing a necessity for a large scale facility investigations. In such samples, an around the mid-gap position of the Fermi level is expected [12] , what results in a highly resistive or even the insulating character. It should be strongly underlined here that the magnetic characterization appears as the easy available and a relatively reliable assessment tool to evaluate the transport properties of (Ga,Mn)N deposited on conducting substrates.

The magnetic characterization has been performed using a Quantum Design MPMS XL Superconducting Quantum Interference Device (SQUID) magnetometer. Because of a low Mn content and a small thickness of the layer deposited on a bulky sapphire substrate, the accurate determination of magnetization as a function of temperature *T* and magnetic field *H* has required an extension of the previous experimental methodology [38][39] to an active, *in situ*, compensation of the dominating flux of the sapphire [40].

Figure 1 summarizes the main magnetic features of the layer studied here. The vanishing of the remnant moment marks the Curie transition temperature $T_C \cong 8$ K, well in the accordance with the data given in ref. [29]. The low-*T* ferromagnetic state of the layer is further corroborated by a well-developed hysteresis curve at *T* = 1.8 K, presented in the inset. Importantly, the TRM signal trails the origin at $T > T_C$ and together with the featureless, flat response at 300 K (inset to Fig. 1) decisively proves the absence of ferromagnetic Mn-rich precipitates or nano-inclusions. Combined, these findings indicate: (i) a homogenous magnetic constitution of the material, (ii) the superexchange mechanism at work, (iii) the Fermi level pinning at the mid-gap $Mn^{3+}$ level, and consequently its highly insulating character. Other works also show that $Mn^{+3}$ or $Mn^{2+}$ ions create relatively deep level defects in the band gap of GaN [41]. Importantly, it should be added that the established a very good command over the MBE growth allows to produce on demand high structural quality (Ga,Mn)N with $x$ up to 10% [10][29]. This is in contrast to other transition metals, such as Fe [42] and Cr [43], which tend to aggregate into metal-rich nanocrystals already for $x > 0.5\%$ [44][45], what results in destroying their carrier trapping capabilities [44].

Such heteroepitaxially grown layers exhibit relatively smooth surface morphology (rms $\cong$ 2 nm, Fig. 2a) with clearly resolved monolayer steps, as indicated by atomic force microscopy (AFM) studies. These terraces exemplified in panels b and c of Fig. 2 are separated by approximately 3 Å high steps. Such a step height compares reasonably well with the 2.6 Å step height expected for one (0001) GaN monolayer.

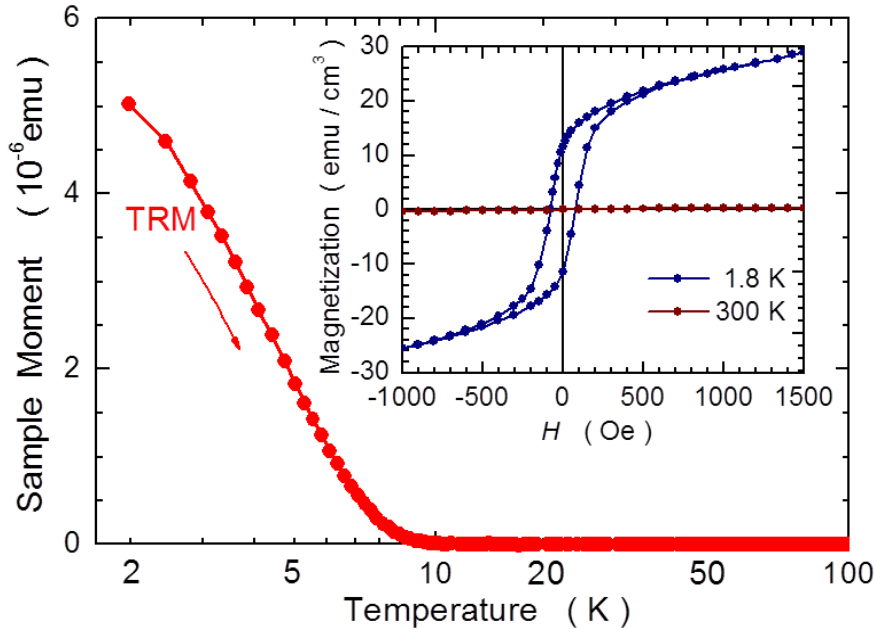

**Fig. 1**. Temperature dependence of the remnant moment (TRM) of the layer investigated here (red bullets). Prior to the TRM measurement the sample had been cooled down at 1 kOe to the base temperature and the field was quenched. The vanishing point of the TRM signal indicates the Curie transition temperature, here about 8 K. Inset: low field hystereses curves at 1.8 and 300 K. The magnetic flux of the sapphire substrate has been removed using *in situ* active compensation method [40].

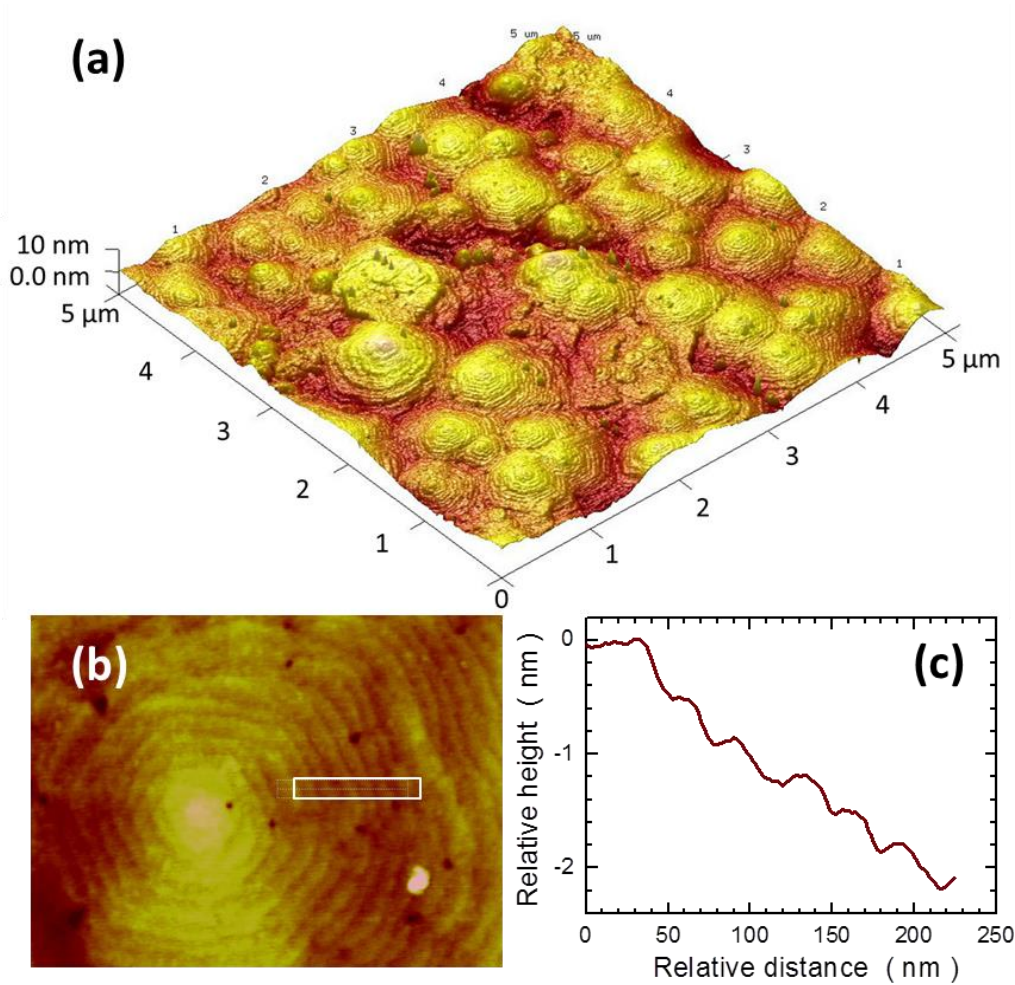

**Fig. 2**. Atomic force microscopy images of the top surface of the as grown sample, i.e. the (Ga,Mn)N layer. (a) 5 × 5 μm$^2$ birds eye view. The root mean square of this image is ~2 nm. (b) Zoom on one of the hillocks, and (c) the linear surface scan documenting well developed terraces separated by approximately 3 Å high steps.

In these growth conditions dislocations containing a screw component contribute to the formation of spiral hillocks [46][47], clearly resolved in our sample by the AFM technique. As shown by Hsu *et al.* [48], the hillocks' tops on the surface of GaN layers are the only places which conduct the current, so we can estimate from Fig. 2a that the density of the conductive dislocations in our structure does not exceed $10^8$ cm$^{-2}$. This is a rather typical density of screw and mixed dislocations in good heteroepitaxial GaN structures.

Two contradicting requirements have to be fulfilled by the investigated device to facilitate the studies. On the one hand side, the junction area should be as small as possible to minimize the number of conductive dislocations, but on the other it should be large enough to permit a relatively easy electrical contact placing. In order to satisfy both conditions we employ a concept elaborated previously by some of the present authors to block the conductive dislocations in vertical heteroepitaxial GaN/(Ga,Mn)N structures by covering their top surface by insulating (dielectric) oxide [17]. As established previously atomic layer deposition (ALD) deposited $Al_2O_3$, as well as $HfO_2$, makes excellent insulating layers on GaN, withstanding up to at least 3 MV/cm at 77 K [49][50]. This simple and very effective solution allowed to demonstrate the existence and to quantify the effectiveness of the piezoelectro-magnetic coupling in (Ga,Mn)N [17], and confirmed its outstanding insulating properties [13][14][15][16], which constitute the functional basis of our device.

The schematic drawing of our device is presented in Fig. 3. $50 \times 50$ μm$^2$ junction to (Ga,Mn)N is defined by the lift-off technique in about 100 nm thick layer of insulating $Al_2O_3$, deposited uniformly on the top (Ga,Mn)N surface by ALD. Next, a macroscopically large Ti(~5 nm)/Au(~150 nm) contact pad (~$200 \times 200$ μm$^2$) is evaporated centrally above the opening in the $Al_2O_3$. The metallization thickness exceeds that of $Al_2O_3$ to assure that the metallic layer connects the top of the (Ga,Mn)N layer with the remaining metallic contact pad resting on the $Al_2O_3$. Here, we underline a double role which (Ga,Mn)N layer plays in this device. While being the subject of the investigation, its highly insulating character renders the lateral conductivity inefficient and so precludes shortening of the device by those conductive dislocations which pierce the (Ga,Mn)N away from the window in $Al_2O_3$. Essentially, this concept is equivalent to an etching a small diameter mesa down to GaN:Si buffer, but eliminates a reactive ion etching step and a rather cumbersome selective electrical contacting of the top of a small mesa.

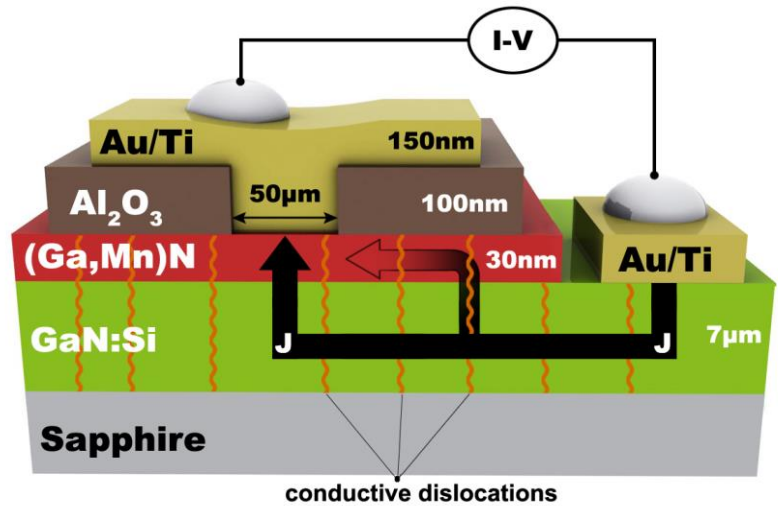

**Fig. 3**. Schematic representation of the device used in this study. The $50 \times 50$ μm$^2$ vertical conductive channel is defined by an opening in deposited by the atomic layer deposition method 100 nm thick insulating $Al_2O_3$ layer (brown). The layer serves at the same time as a base for the top electrical contact. The 30 nm thick (Ga,Mn)N layer (red) is grown by MBE on a reduced threading dislocation density templated sapphire. This 7 μm thick GaN:Si template provides also the bottom electrical connection to the structure. The thick black arrow marks the effective current path, since the highly insulating character of (Ga,Mn)N renders the lateral conductivity inefficient and so precludes shortening of the device by other conductive dislocations.

Three electrical contacts to the bottom n-GaN layer are made during the same Ti/Au metallization process, and are located in the rim part of the substrate. This about 2 mm wide edge strip of the substrate is not overgrown by (Ga,Mn)N since it is obscured by a molybdenum substrate holder during the MBE growth [29]. This approach again eliminates the need for opening an access to the bottom n-GaN layer by reacting ion etching of the top (Ga,Mn)N one, further simplifying the fabrication process. Finally, after placing the sample on the typical dual-in-line integrated circuit socket sample holder, the electrical connection are made by standard bonding using 15 μm Al wire. The electrical measurements are performed in a standard continuous flow cryostat. The current-voltage (I-V) characteristics of the devices in the standard two-terminal configuration are obtained using a "differential conductance mode" of DC Keithley 2182A/6221 tandem (nanovoltmeter/current source).

## 3. Results and discussion

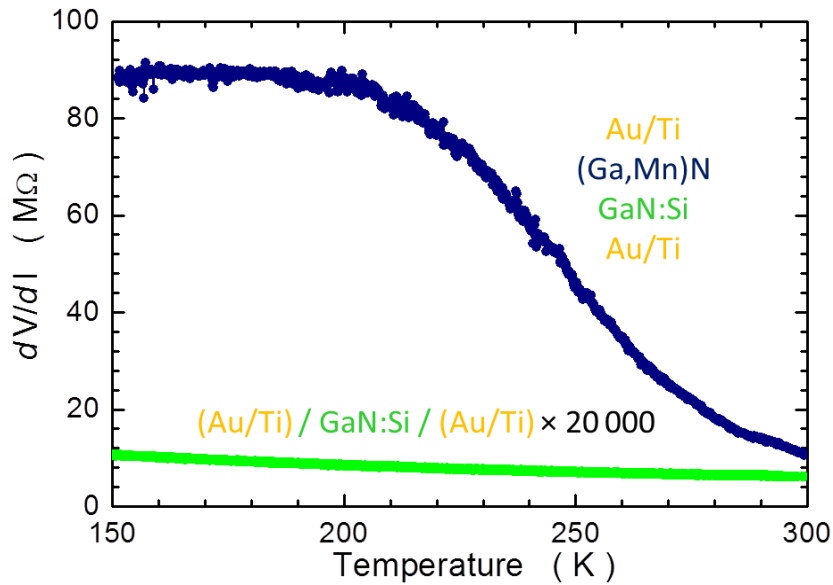

**Fig. 4**. Comparison of the temperature dependence of the two terminal resistance of (navy) the (Au/Ti) / (Ga,Mn)N / GaN:Si vertical device, schematically shown in Fig. 3, and (green) of the GaN:Si buffer. Note that the latter data are multiplied 20000 times. All Au/Ti contacts are formed in the same evaporation/lift off process. The measurements are performed at zero bias condition.

While the two-terminal resistance between any two side contacts to n-GaN remains between 200-500 Ω in the whole studies temperature range, the resistance through the (Ga,Mn)N junctions exceeds 10 MΩ already at room temperature. This resistance further increases on lowering temperature reaching nearly 100 MΩ below 200 K, as presented by navy bullets in Fig. 4. However, already below some 250 K the resistance starts to saturate and the experimental noise increases. We assign this behavior to a signature of a presence of some sparse weakly conducting dislocations which residual parallel conductance starts to determine the junction resistance above some 50 MΩ. Nevertheless, this result constitutes a remarkable progress in comparison to results obtained for similar structures grown on plain *c*-sapphire substrates for which resistances below 1 kΩ were observed [21]. On the other hand, the values presented in Fig. 4 are rather unprecise, since the larger is the resistance of the device the more nonlinear it becomes. This is evidenced in Fig. 5, where we collect the current-voltage (I-V) characteristics of the devices obtained between 150 and 300 K. The figure indicates an asymmetric I-V characteristic as for a good Schottky junction (SJ). No hysteresis, during voltage sweeps, even at the lowest $T$ are observed (not shown), what rules out an influence of deep level defects. We take therefore these findings as a strong indication that the device is nearly devoid of conductive dislocations and that the electron transport at room temperature is governed by the intrinsic electrical properties of the metal/(Ga,Mn)N/GaN:Si stack.

In general, it is an interesting finding that in the material in which on average 2500 conductive dislocations are expected in a $50 \times 50$ μm$^2$ contact window, their influence on electrical transport is barely seen at room temperature. We postulate therefore that some of the dislocations lose their conductive character when they are about to propagate through Mn enriched GaN. And it could be Mn, owing to its mid-gap position of the 2+/3+ level, which would passivate conductive dislocations in the GaN host, an effect similar to that brought about by interstitial O, as already observed experimentally [51] and accounted numerically [52]. The thickness of (Ga,Mn)N in our test structure is rather a moderate one (30 nm), it therefore remains to be seen whether the increase of the defect tolerance in GaN, due to blanking of the conductive dislocations by Mn doping, could become more effective in thicker or more Mn concentrated (Ga,Mn)N layers. Perhaps temperature agitated Mn migration processes occurring both during the growth and/or during the post-growth cooling at the growth chamber play an important role here [29][53].

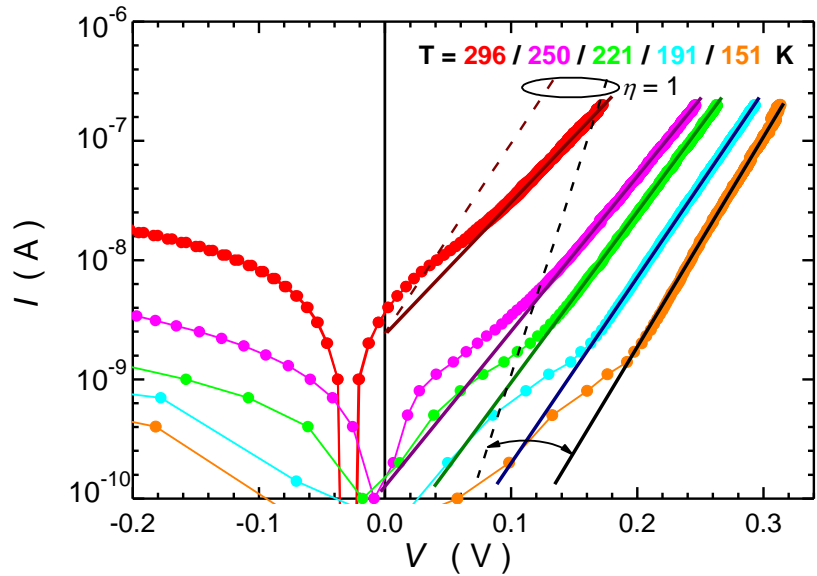

**Fig. 5.** Semi-log plot of I-V characteristics (symbols) at selected temperatures between 150 and 300 K for the studied device. Solid lines: a linear interpolation at forward bias region. Dashed lines represent expected behavior of a Schottky junction in thermionic emission approach at 151 and 300 K (Eq. 1) with the ideality factor $\eta = 1$.

Following the data collected in Fig. 5 we assess an effective metal-(Ga,Mn)N barrier height $\Phi_B$, and the temperature dependence of the fidelity of the SJ. It is clearly seen that logI-V plots are linear over a decade wide range of currents and that they shift to the higher voltages on lowering $T$. In this case we can apply the time honored thermionic emission model [54], in which the forward bias I-V in the region $V > 3k_BT/q$ (e.g. $V > 0.1$ V at 300 K), and in the absence of a series resistance assumes the following form:

$$I = I_S \exp\left(\frac{qV}{\eta k_B T}\right) = AA^*T^2 \exp\left(-\frac{\Phi_B}{k_B T}\right)\exp\left(\frac{qV}{\eta k_B T}\right). \tag{1}$$

Here, $A$ is the contact area, $T$ is the absolute temperature, $q$ is the electron charge, $k_B$ is the Boltzmann constant, $\eta$ is the ideality factor describing how far the junction slope differs from an ideal one. We take the theoretical value of the Richardson constant $A^* \cong 27$ Acm$^{-2}$K$^{-2}$ – corresponding to electron effective mass in GaN: $m^* = 0.22\, m_0$ [55]. The magnitude of the reverse saturation current $I_S$ and $\eta$ are readily obtained from the linear interpolation of semi-log I-V plots at their relevant ranges. These lines are presented in Fig. 5. According to Eq. 1, $\eta$ is inversely proportional to the slope of the interpolating line, whereas its intercept at $V = 0$ yields $I_S$. Then at any given temperature $\Phi_B$ can be readily obtained from:

$$\Phi_B = -k_B T \ln\left(\frac{I_S}{AA^*T^2}\right). \tag{2}$$

The established magnitudes of $\Phi_B$ and $\eta$ are plotted in Fig. 6. At room temperature $\Phi_B = 0.61$ V, which is less than typically reported for metallic SJ to GaN: $0.7 \leq \Phi_B \leq 1$ V [48][56][57]. However, this value stays very close to that found for Ti/GaN SJ, $\Phi_B = 0.59$ V [58], and in our metallization process Ti is deposited first on (Ga,Mn)N to facilitate good adhesion of the main Au contact layer. Moreover, the whole $\Phi_B(T)$ dependency resembles well that established for Ni/n-GaN SJ [59][60]. For example $\Phi_B(150$ K$) \cong 0.4$ eV in all these cases, despite different work functions for both transition metals

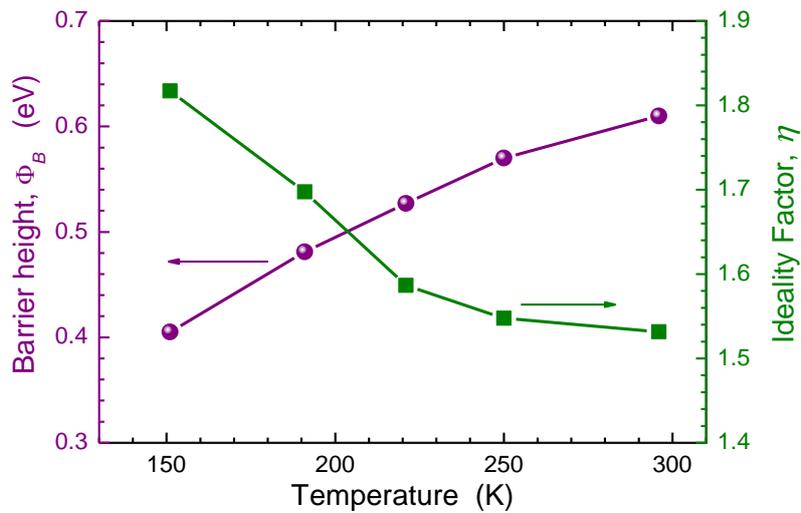

**Fig. 6**. Temperature dependence of the effective Schottky barrier height and the ideality factor $\eta$ established from the forward bias I-V characteristics in thermionic emission approach.

As shown in Fig. 6 the magnitudes of the ideality factor found here are larger than 1. Even at $T \rightarrow 300$ K, $\eta \rightarrow 1.5$, contrary to the prevailing majority of studies, where typically $\eta \cong 1$ for nitride-based junctions at elevated temperatures [57][59][61]. Interestingly, there are reports in which ideality factors of metal / GaN junctions are similar to these found in our study [60][62].

These two findings, temperature dependent magnitude of $\Phi_B$ and $\eta > 1$, indicate a fairly non-ideal nature of the SJ realized in the studied device, or that the transport mechanism across the whole two-terminal device deviates from the thermionic emission theory. For example alloy fluctuations inherent to low-temperature III-V alloys [63], or a non-ideal junction interface due to both intrinsic and extrinsic effects related to preparation or details of metallization were invoked to give an account for a detrimental lateral $\Phi_B$ height inhomogeneity [61][64][65][66]. Other possible effects include a presence of edge or screw dislocations [61][63][67], which introduce high density trap states within the band-gap of the semiconductor what enhances the tunneling probability [23][48][67][68][69][70].

It is worth to underline the fact that in terms of Eqs. 1 and 2 a requirement for $\eta \gg 1$ to reproduce forward bias I-V behavior means that for a given $\Phi_B$ (i.e. for a fixed $I_S$) a smaller current is flowing at the relevant $T$ through our junction than it would if the junction transmission had been determined exclusively by the thermionic emission process. Exemplary dashed lines of slopes corresponding to $\eta = 1$ and giving the same magnitudes of $I_S$ at 296 and 151 K are added to Fig. 5. Following these lines, one can establish that the current measured at +0.2 V is about 15 and 1100 times less than the expected one for the ideal SJ at 296 and 151 K, respectively. It can be understood that such a readily increasing series resistance on lowering temperature is provided in our structure by the deliberately introduced here (Ga,Mn)N barrier. It remains to be experimentally verified which of the effects mentioned above plays indeed a decisive role. Such a finding would shed an additional light on physical mechanisms governing electrical properties of other vertical structures like HEMTs [25][26].

## 4. Conclusions

Vertical transport device has been fabricated from heteroepitaxially grown (Ga,Mn)N on commercially available low threading dislocation density GaN:Si template. The investigated device has been defined lithographically by the opening in the additional insulating dielectric $Al_2O_3$ layer grown by atomic layer deposition on top of (Ga,Mn)N to facilitate electrical bonding. The observed two terminal resistance in the range of tens of M$\Omega$ indicates that a nearly conductive-dislocation-free electrical properties are achieved. A positive role of Mn in electrical blanking of the conductive dislocations has been postulated. The analysis of temperature dependence of the forward bias I-V characteristics in the frame of the thermionic emission model yields Ti-(Ga,Mn)N Schottky barrier height to be slightly lower but close in character to other metal/GaN junctions. However, the large magnitudes of the ideality factor, though not so uncommon in nitride Schottky junctions, point to a sizable blocking of the current in the structure. It still remains to be verified whether it is due to the presence of the insulating (Ga,Mn)N or some other factors which reduce the effective area of the junction are in force.

# Acknowledgments

This study has been supported by the National Science Centre, Poland through FUGA (DEC - 2014/12/S/ST3/00549) and OPUS (DEC - 2013/09/B/ST3/04175) projects.